\documentclass[%
 aip,
 amsmath,amssymb,
 reprint,%
]{revtex4-1}

\usepackage{graphicx}
\usepackage{dcolumn}
\usepackage{bm}

\usepackage[utf8]{inputenc}
\usepackage[T1]{fontenc}
\usepackage{mathptmx}

\begin{document}

\preprint{AIP/123-QED}

\title{Transfer learning driven design optimization for inertial confinement fusion}

\author{K. D. Humbird}
 \email{humbird1@llnl.gov}
\affiliation{ 
Lawrence Livermore National Laboratory
}

\author{J. L. Peterson}
\affiliation{ 
Lawrence Livermore National Laboratory
}

\date{\today}

\begin{abstract}
Transfer learning is a promising approach to creating predictive models that incorporate simulation and experimental data into a common framework. In this technique, a neural network is first trained on a large database of simulations, then partially retrained on sparse sets of experimental data to adjust predictions to be more consistent with reality. Previously, this technique has been used to create predictive models of Omega~\cite{humbirdTL} and NIF~\cite{humbirdcogsim,kustowskiTL} inertial confinement fusion (ICF) experiments that are more accurate than simulations alone. 

In this work, we conduct a transfer learning driven hypothetical ICF campaign in which the goal is to maximize experimental neutron yield via Bayesian optimization. The transfer learning model achieves yields within 5\% of the maximum achievable yield in a modest-sized design space in fewer than 20 experiments. Furthermore, we demonstrate that this method is more efficient at optimizing designs than traditional model calibration techniques commonly employed in ICF design. Such an approach to ICF design could enable robust optimization of experimental performance under uncertainty.

\end{abstract}

\maketitle

\section{\label{sec:intro} Introduction}
Machine learning is gaining momentum in powering science discovery for fusion research. In recent years, Tri-Alpha Energy has used machine learning driven control systems to reach new plasma confinement records at their facility~\cite{trialpha}, inertial fusion researchers have used data-driven models to triple the neutron yield produced at the Omega Laser Facility~\cite{lle_nature}, and machine learning continues to be a popular tool for predicting disruptions in magnetic confinement experiments~\cite{mfml1, mfml2, mfml3}. At Lawrence Livermore National Laboratory machine learning is improving understanding of implosion performance at the National Ignition Facility (NIF)~\cite{JimBayesian}. 

In this work, we propose a machine learning-driven framework for optimizing performance of indirect drive ICF experiments~\cite{Lindl,Atzeni}. The framework utilizes a technique called ``transfer learning’’~\cite{TL,TL1,TL2} to create a neural network model of the experimental performance iteratively, resulting in a model which improves with each experiment. In this hypothetical ICF campaign, a computer model of the experiment is used to demonstrate the methodology, and show that maximum performance can be reached twice as efficiently as more traditional design techniques.

Section \ref{sec:nn} gives a brief introduction to transfer learning and its use in inertial confinement fusion modeling. In section \ref{sec:tlicf} we discuss the proposed technique for a data-driven experimental campaign that seeks to maximize performance by optimizing under uncertainty using two techniques: transfer learning and drive multiplier calibration. Results presented in section \ref{sec:res} show that a transfer learning driven campaign leads to optimized performance more efficiently than traditional drive calibration techniques, which are often unable to reach peak performance in a modest number of experiments.

\section{\label{sec:nn} Transfer learning}
Transfer learning is a popular machine learning technique that involves taking a neural network trained to solve one task, and partially retraining it to solve a different, related task, often with limited data~\cite{TL,TL1,TL2}. Neural network models that solve complex tasks have millions of free parameters that must be tuned by providing the model large quantities of data. Often in scientific applications, there is insufficient experimental data to train these models from scratch, thus scientists can leverage large databases that solve a related task to essentially ``warm start’’ the training of the neural network. A common example is to use computer simulations of the experiments, or theoretical models. These models are often inexpensive compared to the experiment, and can be evaluated thousands of times to create a database for neural network training. 
Many of the free parameters are learned by training on the simulation dataset and then are modified, or ``fine-tuned’’, by passing the real experimental data through the model and allowing a limited number of free parameters to adjust such that the predictions become consistent with the experimental data. 

Transfer learning has shown great success as a method for creating predictive models of ICF experiments~\cite{humbirdTL,kustowskiTL,humbirdcogsim}. We refer the readers to this prior work for details on transfer learning for ICF, and focus this manuscript on the application of this technique to guide a hypothetical experimental campaign. 

Possessing a good model to guide ICF experimental design choices is vital; experiments are expensive, complex, and limited in quantity. Computer models of the experiment are used to search broad design spaces for configurations which lead to desired experimental results, such as maximizing neutron yield production -- an important measure of implosion performance. Computer-based design is only effective if the model is a sufficiently accurate depiction of reality, otherwise the computer model's optimal design might differ from the experimental optimal.  

 Modern ICF simulations, while a good approximation to experimental reality, are known to have sources of error that lead to discrepancies between simulation predictions and experimental measurements in some regions of design space. The simulation can be thought of as providing a ``map’’ of the space – illustrating how neutron yield changes as different design choices are varied (laser pulse shape, capsule size, etc). If the simulation map is incorrect, we can spend a large number of experiments trying to find a performance peak that does not actually exist in the experimental ``map’’. Transfer learning offers a way to update the initially simulation-based map using experimental data, gradually creating a map that looks more like reality, and thus can lead to real performance peaks.
 
In the next section, how transfer learning can be used to guide an ICF experimental campaign is demonstrated with computer-based ``experiments''. The transfer learning approach is contrasted to a traditional calibration method, in which uncertain simulation input parameters are inferred using prior experimental data, and are then used in subsequent simulations to predict the outcome of the next experiment.

\section{\label{sec:tlicf}Data-driven experimental ICF campaign}
To explore the idea of a data-driven experimental campaign, we will perform a thought experiment – treating random perturbations of an ICF multiphysics code as the ``experiment’’ and using the baseline code as our simulation model. This enables optimization to be performed with an unlimited number of experiments,  and enables us to vary the experimental reality by creating several random perturbations of the ICF code in order to get a sense of how, on average, each technique will perform for a variety of possible realities~\cite{nicreview}.

\subsection{Experiment Model}
To determine how efficiently a transfer learned neural network can learn the discrepancy between the simulation model and experimental performance, an ``experimental'' model is generated, and can be evaluated as many times as necessary to optimize performance (unlike real experiments, which cost a significant amount of resources to execute). This ``experiment'' is essentially a random modification to the code Hydra~\cite{hydra}. Specifically, the experimental neutron yield, $Y_{exp}$, is taken to be of the form: 

\begin{equation}\label{eq:hx}
Y_{exp} = A \cdot Hydra\left(Bx \right) + \epsilon
\end{equation}

where $x$ is the input vector, and $A$ and $B$ are matrices. $\epsilon$ is a normally distributed random error representing realistic experimental noise, with a mean zero and standard deviation of 5\% of the experimental yield. Rather than running Hydra itself with a variety of distortions to the inputs, we use a neural network trained on a database of Hydra simulations as an emulator. The neural network leveraged in this works is referred to as a ``deep jointly-informed neural network'' (DJINN)~\cite{djinn}. DJINN is an architecture and weight initialization selection algorithm for easily training deep neural networks and reduces the need to hand-tune network hyper-parameters; it has been shown to perform extremely well on ICF data. 

Integrated hohlraum simulations are computationally expensive, requiring several days on several nodes per simulation to complete. In order to reduce the computational burden of this study, a single database of 2000 simulations is generated with the code Hydra, in which four drive parameters are varied. 
The simulation database used is centered around the NIF experiment N170601, one of the first designs to produce over 10$^16$ neutrons; the model is described in ~\cite{hdc,hdc2}. 
The four inputs that are varied include a cone fraction value (the ratio of laser power in the inner cone of beams to the total power) during the rise to peak power (from a value of 0.25 to 0.4) and at two points in time during the peak of the pulse (from values of 0.25 to 0.4, and 0.05 to 0.5), and an adjustment to the length of the peak of the pulse (from -0.675 ns to 1.25 ns), which adds or removes laser energy. Adjusting the pulse could result in laser energies and powers outside the bounds of what the National Ignition Facility (NIF)~\cite{nif2,NIFMiller} can achieve (480 TW and 1.9 MJ), thus any regions of design space that exceeds these limits are excluded. A DJINN model is trained on the database to create an accurate emulator of Hydra. The experimental model, described by Equation~\ref{eq:hx} is thus an evaluation of a neural network with modified values to the drive inputs.

\subsection{Drive Multiplier Calibration}\label{sec:dm}
Indirect drive ICF simulations are often not consistent with experimental reality when evaluated with as-shot laser pulses and experimental parameters. There are a variety of known and uncertain sources of error in the codes, including 2D approximations to a 3D system, low fidelity physics models to reduce computational cost, and pushing designs into regions of parameter space where the models have not been validated. Thus, in order to bring the simulation outputs into consistency with the experiments, researchers typically adjust a standard set of parameters in the simulation. For integrated hohlraum simulations, like those that are used in this study, it is common to apply multipliers to certain drive parameters to match experimental data. This creates an effective laser drive that might be different than what was actually delivered by the laser in the experiment, but makes up for approximations in physics models like opacities, laser-plasma interactions, laser entrance hole closure rates, etc~\cite{hypyd,hf,bubble}. By adjusting the effective drive, it is often possible to better match the experimental data at least locally, and the calibrated drive can then be used to propose changes to the design that are modest deviations to the prior experiment. 

As a baseline to which transfer learning can be compared, we will execute a hypothetical campaign in which drive multipliers are used to adjust the simulation predictions to be more consistent with experiments, and let the corrected drive guide us toward optimal performance. 

In the drive multiplier calibration approach, a multiplier is applied to each of the four inputs and infer the best values for the multipliers with experimental data. The hypothetical campaign begins with the cone fraction multipliers set at 1.0 and the pulse time shift at 0.0. The prediction error for the experiments is calculated, and the four drive parameters are adjusted to minimize the prediction error. If there is more than one experiment, the drive multipliers that minimize the average error across all experiments are found. The drive parameters that result in the best match to experimental data are used to select the next experiment to run via the optimization algorithm detailed in Section \ref{ssec:opt}.

\subsection{Neural network training}
To create transfer learned models, we start with a simulation-based neural network trained on an ensemble of simulations that span the four-parameter design space discussed in Section ~\ref{sec:dm}. An ensemble of 2000 Latin hypercube sampled~\cite{lhs} 2D hohlraum simulations are run using the radiation hydrodynamics code Hydra. Neural networks are trained on the data using the code DJINN, producing models that map from the four drive inputs to the output quantities of interest (QOIs) – the neutron yield of the implosion, as well as the energy of the laser and the peak power of the laser. The latter two quantities are necessary for limiting the choice of possible experiments to those that fall within the operating boundaries of the NIF. 
The DJINN models are each an ensemble of five neural networks with Bayesian dropout employed~\cite{dropout}, such that the networks can be evaluated multiple times (100) to get a distribution of predictions for the QOIs. For power and energy, the median prediction is used to set the operating boundaries of the laser. For neutron yield, the uncertainty in the yield prediction is used in the cost function that is optimized to select new experiments, detailed in the next subsection. The DJINN models all have a maximum depth of five layers, with dropout keep rates of 0.97 in each layer. They are trained for 500 epochs with a batch size of 100 and a learning rate of 0.003. Table~\ref{tab:nnmetrics} summarizes the median predictive capability of each of the network models. 

\begin{table}[]
\caption{Error metrics on holdout validation data for DJINN models.}
\label{tab:nnmetrics}
\begin{tabular}{llll}
       & Explained Variance & Mean Absolute Error & Mean-Squared Error \\
Power  & 0.932              & 3.19                & 25.16              \\
Energy & 0.991              & 0.017               & 0.0008             \\
Yield  & 0.988              & 9.405               & 171.2             
\end{tabular}
\end{table}

After each experiment is executed, the last layer of the models is retrained starting from the baseline simulation-only model. The model is retrained with the available experimental data by adjusting the weights and biases of the last layer, training for 250 epochs with a batch size of one and learning rate of 0.0001. 

\subsection{Optimization}\label{ssec:opt}
In the drive calibration approach, the multipliers that best fit the data are applied to the simulation model. This ``corrected'' simulation model is used to select the next experiment to run in order to maximize a quantity called ``expected improvement’’. Expected improvement is defined by the following equation: 

\begin{equation}
EI = -(z-Y_{max})*normCDF(z)+std*normPDF(z),
\end{equation}

where $Y_{max}$ is the current maximum experiment yield and $z$ = (mean-$Y_{max}$)/std, and the mean and standard deviation are the predictions of yield performance (with uncertainty) from the DJINN model. 

Neither transfer learning nor drive multiplier calibration are reliable at extrapolating, thus to prevent the models from suggesting new experiments far outside the limits of where they are expected to be valid, we only consider a small volume of design space about our current highest yield experiment as candidates for the next experiment. To choose the next experiment, 10,0000 candidate points are created within a Latin hypercube of side length R. The side length takes on one of three values: if the last experiment was more than 1.2x the previous max yield, R is set to 0.25, since it is likely near a high yield peak. If the previous experiment is less than 0.8x the previous max yield, R is set to 0.75 to expand the search space. Otherwise, a standard radius of 0.5 is used (the LHS is traditionally scaled [0,1]). The LHS is scaled to the appropriate cube size, sampled 10,000 times, and each model is used to predict the neutron yields (with uncertainties) for these points. The expected improvement is computed for each of the candidate points, and the point which maximizes expected improvement is selected as the next experiment. The experimental model is evaluated at the selected point, and the resulting outcome is added to the collection of experiments for the next iteration.

\section{Results}\label{sec:res}
The campaigns begin with M number of randomly chosen experiments to provide the models space-filling data, and then we begin running experiments that optimize the expected improvement for the yield according to each of the models under consideration: the transfer learned model, the drive calibration model, and the experimental data only neural network. 

Since the $A$ and $B$ matrices that define the experiment are randomly generated, the thought experiment can be repeated dozens of times to get a statistical picture of how quickly maximum yield can be achieved in a campaign using transfer learning or drive multiplier calibration. 

\begin{figure}
\includegraphics[width=0.45\textwidth]{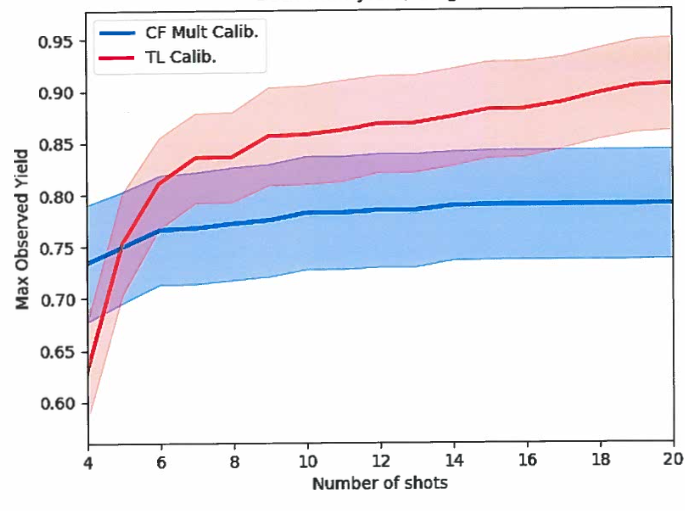} 
\caption{\label{fig:thoughtlineD} The maximum observed yield, normalized by the true maximum achievable yield, as a function of the number of experiments, or shots, executed. In this study, the experimental model is defined by a diagonal matrix $B$. In red is the result of the transfer learned model, in blue is the results of calibrating cone fraction multipliers. The error bounds illustrate the variation in performance as the random matrices are varied. At the low data limit the drive calibration method achieves better performance than the TL model. However, the TL model increases in performance rapidly as data is acquired, ultimately reaching 90\% of the peak yield at 20 experiments, where the calibration method learns more slowly, and reaches only around 80\% of the peak yield.}
\end{figure}

To illustrate the performance of each approach -- transfer learning and drive calibration -- the transformation $B$ is first limited to be a diagonal matrix, such that the drive calibration technique could eventually learn the mapping to within the experimental noise levels. That is, the transformation that maps simulations to experiments is just a specific set of values for the four drive multipliers. The maximum yield obtained by each model as a function of the number of experimental data points acquired is shown in Figure~\ref{fig:thoughtlineD}. 

For diagonal perturbations, the models perform similarly. The drive multiplier inference is more accurate in the very low data limit, but is out-performed by transfer learning once more than six experimental data points are available. 

\begin{figure}
\includegraphics[width=0.5\textwidth]{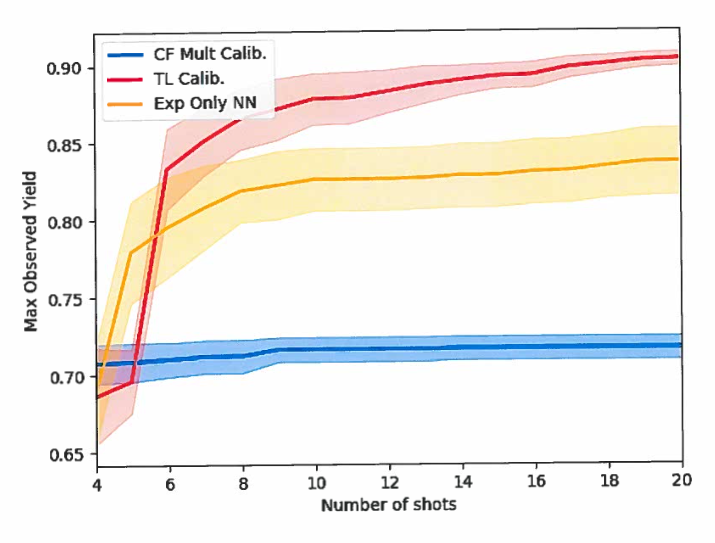} 
\caption{\label{fig:thoughtline} The maximum observed yield, normalized by the true maximum achievable yield, as a function of the number of experiments executed. In this study, the experimental model is defined by a dense random matrix $B$. In red is the result of the transfer learned model, in blue is the results of calibrating drive multipliers, and in yellow is  neural network trained only on experimental data. The error bounds illustrate the variation in performance as the random matrices defining the experimental model are varied. The transfer learned model consistently learns more efficiently than the other techniques, reaching almost 90\% of the true maximum yield after 15 experiments.}
\end{figure}

A natural question with regard to transfer learning is whether or not the warm-start provided by training on large simulation databases is necessary. To address this question, the next study considers a randomly initialized neural network trained exclusively on the experimental data. In the limit of a large quantity of experimental data, the transfer learning process is unnecessary -- a neural network can learn the mapping from inputs to outputs without the initial knowledge gained from the simulation data. However, with too few experiments the neural network is unlikely to learn the mapping without over-fitting to the training data, due to the large number of degrees of freedom in the model and the few data points available. To illustrate the benefits of transfer learning at the low data limit, an experimental data-only neural network is constructed using the DJINN algorithm with a depth of four hidden layers, trained for 500 epochs with a learning rate of 0.004 and a batch size of one.
Each model is tested on the more challenging problem of dense random matrices for $B$ in Eq.~\ref{eq:hx}. The results are illustrated in Figure ~\ref{fig:thoughtline}.

In the case of the more complex transformation, it takes each method significantly more data to begin converging on an optimized yield. In the very low data limited all methods start at similar yields, but the neural networks outperform drive calibration after five experiments. Drive calibration is unable to learn the complex transformation, and plateaus at around 70\% of the true maximum yield. The experimental data only neural network improves the maximum observed yield rapidly in the first few experiments, but plateaus at 85\%, possibly due to over-fitting to the experimental data and not accurately predicting the outcome of proposed experiments. The transfer learned model improves the yield after each experiment, reliably reaching 90\% of the true maximum yield by 20 experiments. 

\begin{figure*}
\includegraphics[width=0.9\textwidth]{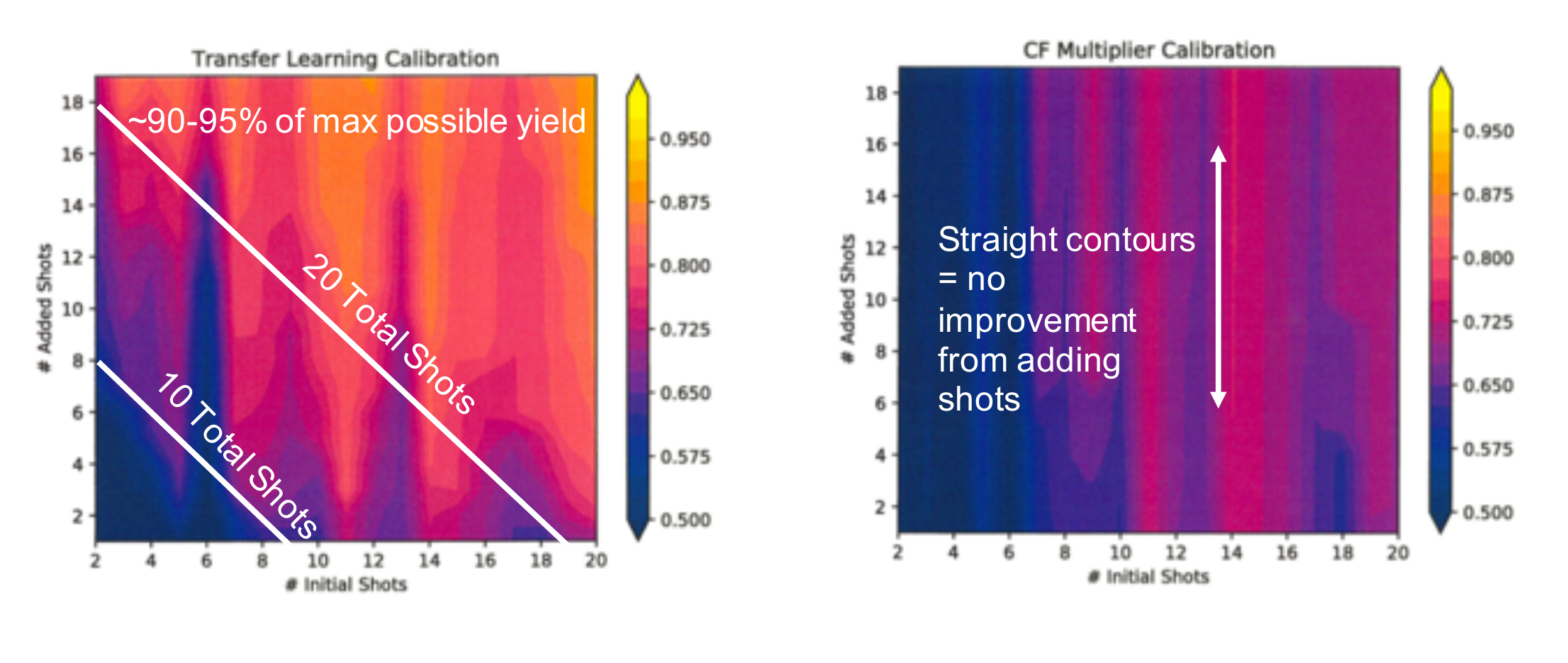}\label{fig:contour}
\caption{\label{fig:nn} The maximum achieved experimental yield as the number of random and optimal experiments are varied for transfer learning and cone fraction multiplier calibration techniques. Transfer learning reaches about 90\% of the maximum possible yield in 20 experiments; cone fraction multiplier calibration often cannot find the maximum yield design.}
\end{figure*}

Finally, we consider the balance of exploration and exploitation in the optimization process with a transfer learned model. The network needs a few initial data points in order to begin learning the ``map'' of the design space. If the sampling method does not explore enough of the space, the models could get stuck in local optima during optimization. Figure \ref{fig:contour} illustrates how quickly transfer learning and drive calibration reach maximum yield as a function of the number of space-filling experiments (M) and the number of experiments that are chosen by optimizing expected improvement (N). At around 20 total experiments the transfer learned model has achieved 95\% of the maximum achievable yield. There does appear to be a weak dependence on the number of initial shots, with the network requiring fewer iterations to reach 80\% of the maximum yield when training on 10 or more space filling designs. However, even with only 2-4 initial experiments, the model does reach 90\% of the maximum yield within 20 total shots. 
In contrast, adjusting drive multipliers plateaus at a maximum of 80\% of the true peak yield independent of the number of experiments initially or iteratively chosen. Drive multiplier calibration assumes the simulation ``map'' of the design space is approximately correct, and only enables you to scale, rotate, or translate the map. If there are off diagonal values in A or B, for example, adjustments to the drive will not recover these discrepancies. Transfer learning, however, is a nonlinear calibration technique that can learn complex adjustments to the response surface given adequate experimental data. Transfer learning is a powerful technique for learning discrepancies between simulation and experimental data when the form of the discrepancy is unknown, or suspected to be complex and difficult to parameterize.  
 
\section{\label{sec:concl}Conclusions}
Transfer learning is a promising method for creating predictive models of ICF experiments, and offers a new approach to experimental design for optimizing performance. In this work, we compare a transfer learning driven hypothetical ICF campaign to one in which drive multipliers are calibrated to experimental data. Transfer learning consistently optimizes designs more efficiently than drive multiplier calibration, often finding yield within 10\% of the maximum achievable yield in the design space in fewer than 20 experiments. Drive calibration, on the other hand, does not always find optimal designs due to its inability to model nonlinear discrepancies between the simulation predictions and experimental reality. This technique is limited to modest dimensional design spaces due to the expense of generating enough simulations on which to train a neural network, however this approach to experimental design could be adapted to use more flexible transfer learning methods ~\cite{humbirdcogsim} to search broad design spaces with experimentally informed models.

\begin{acknowledgments}
This work was performed under the auspices of the U.S. Department of Energy by Lawrence Livermore National Laboratory under Contract DE-AC52-07NA27344. Released as LLNL-JRNL-834445.
This document was prepared as an account of work sponsored by an agency of the United States government. Neither the United States government nor Lawrence Livermore National Security, LLC, nor any of their employees makes any warranty, expressed or implied, or assumes any legal liability or responsibility for the accuracy, completeness, or usefulness of any information, apparatus, product, or process disclosed, or represents that its use would not infringe privately owned rights. Reference herein to any specific commercial product, process, or service by trade name, trademark, manufacturer, or otherwise does not necessarily constitute or imply its endorsement, recommendation, or favoring by the United States government or Lawrence Livermore National Security, LLC. The views and opinions of authors expressed herein do not necessarily state or reflect those of the United States government or Lawrence Livermore National Security, LLC, and shall not be used for advertising or product endorsement purposes. Data subject to third party restrictions. 

\end{acknowledgments}

\bibliography{bib.bib}

\end{document}